\documentstyle[twoside,fleqn,espcrc2_myown]{article}

\renewcommand{\a}{\alpha}
\renewcommand{\b}{\beta}
\renewcommand{\d}{\delta}

\newcommand{\g}{\gamma}

\renewcommand{\l}{\lambda}

\newcommand{\r}{\rho}

\newcommand{\real}{{{\rm I} \kern -.19em {\rm R}}}

\newcommand{\complex}{{{\rm I} \kern -.59em {\rm C}}}
\newcommand{\naturel}{{{\rm I} \kern -.2em {\rm N}}}

\newcommand{\ie}{{{\em i.e.},\ }}
\newcommand{\eg}{{{\em e.g.},\ }}

\newcommand{\be}{\vspace{-1mm}\begin{equation}}
\newcommand{\ee}{\end{equation}\vspace{-1.8mm}}
\newcommand{\een}[1]{\label{#1}
\end{equation}\vspace{-1.8mm}}
\newcommand{\ba}{\vspace{-1mm}\begin{array}}
\newcommand{\ea}{\end{array}\vspace{-1.5mm}}
\newcommand{\beq}{\vspace{-1mm}
\begin{equation}\begin{array}{rcl}}
\newcommand{\eeq}{\end{array}\end{equation}\vspace{-1.5mm}}
\newcommand{\ercln}[1]{\end{array}\label{#1}
\end{equation}\vspace{-1.5mm}}
\newcommand{\nn}{\nonumber}
\renewcommand{\=}{\!\!\!\!&=&\!\!\!\!}

\newcommand{\equ}[1]{(\ref{#1})}
\newcommand{\journal}[4]{{\sl #1~}{\bf #2},\,(19#3)\,#4;}
\newcommand{\hpa}{\journal {Helv. Phys. Acta}}

\newcommand{\np}{\journal {Nucl. Phys.}}

\makeatletter
\@addtoreset{equation}{section}
\makeatother

\newcommand{\AmS}{{\protect\the\textfont2
  A\kern-.1667em\lower.5ex\hbox{M}\kern-.125emS}}

\title{
\vspace{-15mm}
{\normalsize\null July 1996\hfill NEIP-96-003\\
\vspace{-4mm}
\null\hfill hep-ph/9607368\\
}
Finiteness in $N\!\!=\!\!1$ SYM Theories}

\author{C. Lucchesi\address{Institut de Physique,
Universit\'e de Neuch\^atel\\
1 rue Breguet, CH -- 2000 Neuch\^atel (Switzerland)}%
\thanks{Talk given at SUSY96, The 4th 
International Conference on Supersymmetries 
in Physics (College Park, MD, USA), May 29 - 
June 1, 1996. Work supported in part by the Swiss
National Science Foundation.}}

\begin{document}
\thispagestyle{empty}

\begin{abstract}
I present a criterion for all-order finiteness
in $N\!\!=\!\!1$ SYM theories. Three applications
are given; they yield all-order finite
$N\!\!=\!\!1$ SYM models with global symmetries
of the superpotential.
\end{abstract}

\maketitle
\section{INTRODUCTION}\vspace{-2mm}
The aim of this paper is to present applications
of the criterion for all-order finiteness in
$N\!\!=\!\!1$ SYM of \cite{LPS}. All-order
finiteness is here meant in the sense of exact
vanishing of the perturbative $\b$-functions.

The all-order finiteness criterion is an
exact result, with hypotheses operating exclusively at
the one-loop level. It is based on the structure
of the supercurrent anomaly multiplet, which
relates the
conformal anomalies to the  axial ones.
The axial anomalies
being non-renormalized, they are given by their
one-loop values.
Vanishing of the latter is guaranteed by
the hypothesis that the one-loop gauge
$\beta$-function, as well as the one-loop
anomalous dimensions, vanish (these two
conditions are known to yield one-loop
finiteness \cite{PW}).
A further hypothesis on the unicity of
the solution to the conditions of vanishing
one-loop Yukawa $\b$-functions comes from
imposing, as a consistency requirement,
that reduction of the couplings be verified.
Therefore, the all-order finiteness result
is the one-loop result supplemented by a
consistency requirement for higher orders
[hyp. (iv) below].
We shall first review one-loop finiteness
(section \ref{sec2}). In section \ref{sec3},
we state the all-order result. Section
\ref{sec4} presents three applications.
For related approaches to all-order
finiteness in $N=1$ SYM, see
\cite{ermushev,DJZ,strassler}.
\vspace{-2.2mm}
\section{ONE-LOOP FINITENESS}\label{sec2}
\vspace{-2mm}
Consider an $N\!\!=\!\!1$ super-Yang-Mills
theory with simple gauge group (gauge coupling
$g$) and  superpotential
$W(\phi^i)= \frac{1}{2}\,m_{ij}
\,\phi^{i}\,\phi^{j}+
\frac{1}{6}\,\l_{ijk}\,\phi^{i}
\,\phi^{j}\,\phi^{k}$. The matter fields
$\phi^{i}$ transform
under the irrep. $R_{i}$. The one-loop gauge
$\beta$-function is given by
\be
\beta^{\scriptscriptstyle\, (1)}_{g}=
\frac{g^3}{16\pi^2}\,\bigl[\,\sum_i\,
T(R_{i})-3\,C_{2}(G)\,\bigr].
\label{betag}
\ee
The one-loop Yukawa $\beta$-functions
\be
\beta^{\scriptscriptstyle\, (1)}_{ijk} =
\l_{ijl}\,\gamma^{l\ \scriptscriptstyle\, (1)}_{\ k}+
 \l_{ikl}\,\gamma^{l\ \scriptscriptstyle\, (1)}_{\ j}+
 \l_{jkl}\,\gamma^{l\ \scriptscriptstyle\, (1)}_{\ i}
\ee
are combinations of the one-loop anomalous dimensions
\be
\gamma^{i\ \scriptscriptstyle\, (1)}_{\ j}
=\frac{1}{32\pi^2}\,[\,
\l^{ikl}\,\l_{jkl}-2\,g^2\,C_{2}(R_{i})
\delta^i_{j}\,].
\label{gamay}
\ee
Necessary and sufficient conditions for one-
(and two-) loop finiteness \cite{PW}
result from demanding that \equ{betag}, resp.
\equ{gamay}, vanish, \ie
\be
\sum_i T(R_i) = 3\, C_2(G),\label{1st}
\ee
\vspace{-2mm}
\be
\l^{ikl} \l_{jkl} = 2\,\, g^2 \, C_2(R_i)\,
\delta ^i_j .
\label{2nd}
\ee
\vspace{-2.2mm}
\section{ALL-ORDER FINITENESS}\label{sec3}
\vspace{-2mm}
At all orders, finiteness is guaranteed
by the following criterion
\cite{LPS}: {\bf if} {\bf (i)} the gauge
anomaly vanishes,
{\bf (ii)} $\b_g^{\scriptscriptstyle\, (1)}\!=\!0$,
{\bf (iii)} the conditions
$\g^{i\ \scriptscriptstyle\, (1)}_{\ j} \!=\!0$
possess solutions of the
form $\l_{ijk} =\r_{ijk}\,g,\ \r_{ijk}
\in\complex$, and
{\bf (iv)} these solutions of
$\g^{i\ \scriptscriptstyle\, (1)}_{\ j} \!=\!0$
are isolated and non-degenerate\footnote{By
``isolated", we
mean that the zeroes cannot be multiple,
whereas by ``non-degenerate" we forbid parametric
families.} solutions of
$\b^{\scriptscriptstyle\, (1)}_{ijk} \!=\!0$,
{\it {\bf then} {\it each} of these solutions
corresponds to an $N\!\!=\!\!1$ SYM
model with {\it one} independent
coupling constant
(\eg the gauge coupling $g$) which does
not run, \ie $\b_g\!=\!0$ at all orders.}

For the proof, see \cite{LuZou} and the
original literature \cite{LPS,pisi}.
In order to obtain a SYM model with one
isolated and
non-degenerate solution (\ie a unique
solution
for {\it that} model), one generally
needs to restrict
the superpotential by imposing global,
chiral or
discrete, symmetries. It turns out that
one solution of
$\g^{i\ \scriptscriptstyle\, (1)}_{\ j} \!=\!0$
which is isolated and non-degenerate
when regarded as a solution of
$\b^{\scriptscriptstyle\, (1)}_{ijk} \!=\!0$
corresponds, if it exists,
to a given global chiral symmetry of the
superpotential.
Such chiral symmetries are denoted by
\be
\d_{a}\,\phi^i=i\,e_{a\ j}^{\ i}\phi^j,\quad
\d_{a}\,\phi_i^\dagger=-i\phi_j^\dagger\,e_{a\ i}^{\ j},
\label{trou}
\ee
where $e_{a}=e_{a}^\dagger$ are the Hermitean
charges. Symmetry of the superpotential is then
guaranteed provided
\be
\l_{ijl}\, e_{a\ k}^{\ \,l} + \l_{jkl}
\, e_{a\ i}^{\ \,l} +\l_{kil}\, e_{a\ j}^{\ \,l}=0.
\label{46}
\ee
\vspace{-2.2mm}
\section{APPLICATIONS}\label{sec4}
\vspace{-2mm}
\subsection{$\bf SU(6)_{\rm\bf gauge}$
$\bf N\!\!=\!\!1$ SYM finite model}
The first application \cite{LPS} is based on
an $SU(6)_{\rm gauge}$ model which is known
to be one-loop finite \cite{HPS}. It contains
matter supermultiplets in the
${\bf 6},\ \overline{\bf 6},\ \overline{\bf 15}$
and ${\bf 21}$
with multiplicities $(8,16,1,1)$, and has been
chosen for purely illustrative purposes. The
superpotential writes (repr. indices suppressed)
\be
W \! = \! \l_{(ab)}\,{\bf\overline 6}^a
{\bf\overline 6}^b {\bf 21}
+ \l^{[ij]}\, {\bf 6}_i {\bf 6}_j {\bf\overline{15}}
+\l_3\,{\bf\overline{15}}^3,
\ee
where $\l_{(ab)}$, $\l^{[ij]}$ and $\l_3$ are
the Yukawa couplings, $i,\!j\!=\!1...8$,
$a\!\,,\!b\!=\!1...16$.
The $\gamma^{\scriptscriptstyle\, (1)}\!=\!0$
conditions \equ{2nd} yield
\beq
\l^{(ac)}\l_{(cb)}\=\a\,g^2\,\d^a_b,\nn\\[1mm]
\l^{(ac)}\l_{(ca)}\=16\a\,g^2,\nn\\[1mm]
4\,\l_{[ik]}\l^{[kj]}\=7\a\,g^2\,\d^j_i,\nn\\[1mm]
2\,\l_{[ik]}\l^{[ki]}+9\,|\l_3|^2\=28\a\,g^2,\nn
\label{gaze}
\eeq
with $\a$ some constant. The $\l_{(ab)}$'s
and $\l^{[ij]}$'s obeying \equ{gaze} form
parametric families of zeroes, and $\l_3$ is a
degenerate (double) zero. These solutions
are hence neither isolated nor non-degenerate
solutions of
$\b^{\scriptscriptstyle\, (1)}_{ijk} \!=\!0$,
and hypothesis (iv) of the finiteness
criterion is not satisfied.

Let us then pick one of the above solutions,
denote it by $(l_{(ab)},k^{[ij]})$, and
replace it into the superpotential. The latter,
restricted superpotential, is invariant under
chiral transformations of the type \equ{trou},
\ie
\be
\d_{(l)}{\bf\overline 6}^a=i\,e_{(l)\, b}^{\ a}
{\bf\overline 6}^b,\qquad
\d_{(k)}{\bf 6}_i=i\,e_{(k)\, i}^{\ j}{\bf 6}_j
\label{holo},
\ee
provided relations of the form \equ{46} among
generators and Yukawa couplings
\beq
l_{(ac)}e_{(l)\,b}^{\ c}+
l_{(bc)}e_{(l)\,a}^{\ c}\= 0,\\[1.5mm]
k^{[il]}e_{(k)\,l}^{\ j}+
k^{[jl]}e_{(k)\,l}^{\ i}\=0,\label{hola}
\eeq
hold. Considering now the Yukawa couplings
as metrics, we infer that the $e_{(l)}$'s
generate the group preserving $16\!\times\! 16$
bilinear symmetric metrics, \ie the orthogonal
group $O(16)$, whereas the $e_{(k)}$'s generate
the group preserving $8\!\times\! 8$ bilinear
antisymmetric metrics, \ie the symplectic group
$Sp(8)$. Other symmetries (see \cite{LPS}) make
$\l_3$ an isolated zero.
We have thus obtained an all-order finite
$N\!\!=\!\!1$ SYM model with symmetry group
$SU(6)_{\rm gauge}\!\times\! [O(16)
\!\times\! Sp(8)]_{\rm global}$.
\vspace{-1mm}
\subsection{$\bf SU(5)_{\rm\bf gauge}$
$\bf N\!\!=\!\!1$ SYM finite model}
We consider now a more phenomenologically-minded
application of all-order finiteness (see
\cite{LuZou} and references therein). Among
the $SU(5)$ one-loop finite models of
\cite{HPS}, only one allows for three fermion
generations and adjoint matter. It is built
out of the
chiral supermutiplets
${\bf 5},\,\overline{\bf 5},\,{\bf 10},\,
\overline{\bf 10}$ and ${\bf 24}$
with multiplicities $(4,7,3,0,1)$.
The superpotential (repr. indices suppressed) is
\beq
W \={g_{ija}\over 2}{\bf 10}_i {\bf 10} _j H_a
+ {\bar g}_{ija}{\bf 10}_i {\bar {\bf 5}}_j
{\bar H}_a\nn\\[1mm]
&&\!\!\!\!\!\!+{g^\prime_{ijk}\over 2}{\bf 10}_i
{\bar {\bf 5}}_j {\bar {\bf 5}}_k
+{q_{ia b}\over 2}{\bf 10}_i {\bar H}_a
{\bar H}_b\nn\\[1.5mm]
&&\!\!\!\!\!\!+f_{a b}\ {\bar H}_a {\bf 24} H_b
+h_{i a}\ {\bar {\bf 5}}_i {\bf 24} H_a
+p\ {\bf 24}^3,
\label{superpot2}
\eeq
where $i,\!j,\!k\!=\!1...3$ and
$a,\!b \!=\!1...4$. The ${\bf 10}_i$'s and
${\bar {\bf 5}}_i$'s
are the usual three generations,
and the ${\bf 24}$ contains the scalar
superfield.
The four (${\bf 5}+{\bar {\bf 5}}$) Higgses are
denoted by $H_a$, ${\bar H}_a$.
The $\gamma^{\scriptscriptstyle\, (1)}\!=\!0$
conditions \equ{2nd} can be computed to be
\beq
4  {\bar g}_{ij a}
{\bar g}^{ijb}  \!+\! {24\over 5}
 f_{ac} f^{bc} \!+\!4 q_{iac} q^{ibc}
\={24\over 5}  g^2
  \delta _a ^b ,\\[1mm]
3 g_{ija} g^{ijb}
 \!+\! {24\over 5}  f_{ca} f^{cb}
 \!+\! {24\over 5}
 h_{ia} h^{ib} \={24\over 5}
  g^2  \delta _a ^b ,\\[1mm]
 4  {\bar g}_{ki a}
 {\bar g}^{kja} \!+\!{24\over 5}
 h_{ia} h^{ja} \!+\!4
{g^\prime}_{ikl} {g^\prime}^{jkl}
\={24\over 5}  g^2
\delta _i^j,\\[1mm]
2  {\bar g}_{ik a}{\bar g}^{jka}
 \!+\!3 g_{ika}  g^{jka} \!+\!q_{iab}
 q^{jab} \nn\\
 \!+\!{g^\prime}_{kli}
 {g^\prime}^{klj}
\={36\over 5}
  g^2  \delta _i ^j,\\[1mm]
f_{ab} f^{ab} \!+\!{21\over 5}
  pp^* \!+\! h_{ia} h^{ia}
\= 10 g^2.
\label{yg}
\eeq
Finiteness at all orders is achieved by imposing
to the superpotential
a $Z_{7}\!\times\! Z_{3}$ discrete symmetry with
$Z_7$-charges (1;2;4;4;1;2;5;3;6;0) and $Z_3$-charges
(1;2;0;0;0;0;1;2;0;0) [for the ten fields
$({\bf 10}_i;\bar {\bf 5}_i;H_a)$],
plus a multiplicative $Q$-parity (see \cite{LuZou}).
These symmetries enforce
$g_{iii}^2 \!=\!{8\over 5}g^2\!,\ \bar g_{iii}^2
\!=\!{6\over 5}g^2\!,
\ f_{44}^2 \!=\!g^2\!,\ p^2 \!=\!{15\over 7}g^2\!,$
for $i=1,2,3$, and all other Yukawa couplings vanish.
\vspace{-1mm}
\subsection{SYM finite models with global
$\bf U(1)$'s}
Finiteness can be attained by imposing that
the superpotential be invariant under a product
of global $U(1)$ groups. One specific example
has been successfully worked out \cite{LPnotes}.
Starting from some one-loop finite $N\!\!=\!\!1$
SYM model extracted from the list of \cite{HPS},
one constructs {\it one} explicit solution to
the conditions of vanishing one-loop
$\g$-functions. The latter solution is then
replaced into the superpotential, hence
yielding a restricted superpotential $\hat W$.
The next task is to find a minimal set of
global $U(1)$ symmetries, say $n$ such groups,
that constrain the original superpotential to
be just $\hat W$. The $n$ associated Abelian
charges carried by the matter superfields have
to be chosen accordingly. If this can be done
and yields isolated and non-degenerate solutions
to $\b^{\scriptscriptstyle\, (1)}_{ijk} \!=\!0$,
what one ends up with is an all-order finite
$N\!\!=\!\!1$ SYM model with symmetry group
$G_{\rm gauge}\!\times\!_{i=1}^n U(1)^i_{\rm global}$.
It is not clear whether such an approach can be
used in general, or if it applies to a definite
subset of all lower-orders finite models.
\vspace{-2.2mm}
\section{CONCLUSIONS}\vspace{-2mm}
In the process of constructing
all-order finite
SYM theories, one first reduces
 the number of independent Yukawa couplings
by means of global symmetries. Then one checks
if the solution of
$\g^{i\ \scriptscriptstyle\, (1)}_{\ j} \!=\!0$
considered as a solution of
$\b^{\scriptscriptstyle\, (1)}_{ijk} = 0$ is
isolated and non-degenerate. If not,
the process has
 to be restarted, imposing an enlarged
global symmetry
of the superpotential. The process stops
successfully if
unicity of the solution of
$\b^{\scriptscriptstyle\, (1)}_{ijk} = 0$ is attained.

There can be different,
arbitrarily multiple or degenerate solutions to
$\g^{i\ \scriptscriptstyle\, (1)}_{\ j} \!=\!0$.
Each of them {\it may} yield
a finite SYM model with global symmetries, assuming
that such symmetries exist. If more than one finite
model can be constructed for a given unconstrained
$N\!\!=\!\!1$ SYM theory, then each of these models
corresponds to the original theory with an additional
global symmetry specific to that model.
One may hope that the global symmetries which
are necessary for finiteness turn out to be
physically relevant
and to carry predictive power. Of course, one
can also check for finiteness after imposing
global symmetries that are motivated by phenomenology
(as, \eg family symmetry).

Applications of the finiteness criterion
to SGUTS with $SU(6)$ and $SU(5)$ gauge groups,
and specified matter contents,  have been presented.
The latter models are
shown to be all-order finite
provided one imposes the global Lie-type,
resp. discrete, symmetries
$[O(16)\!\times\! Sp(8)]_{\rm global}$ and
$Z_{7}\!\times\! Z_{3}$.
A more generic example has been outlined,
for which all-order finiteness is achieved
by imposing a product of global $U(1)$
symmetries with suitable charges.

{\bf Acknowledgements:} The author thanks
O. Piguet for discussions and a careful reading
of the manuscript.
\vspace{-2mm}


\begin{thebibliography}{9}
\vspace{-2mm}
\bibitem{LPS} C. Lucchesi, O. Piguet and K. Sibold,
 \hpa{61}{88}{321}
{\sl Phys. Lett.} {\bf B201} (1988) 241;
C. Lucchesi, Proc. of SUSY 95, Paris,
June 15-19, 1995, hep-th/9510078.
\bibitem{PW} A.J. Parkes and P.C. West,
{\sl Phys. Lett.} {\bf B138} (1984) 99;
{\sl Nucl. Phys.} {\bf B256} (1985) 340;
P. West, {\sl Phys. Lett.} {\bf B137} (1984) 371;
D.R.T. Jones and A.J. Parkes, {\sl Phys. Lett.}
{\bf B160} (1985) 267;
D.R.T. Jones and L. Mezincescu,
{\sl Phys. Lett.} {\bf B136} (1984) 242; {\bf B138}
(1984) 293; A.J.~Parkes,
{\sl Phys. Lett.} {\bf B156} (1985) 73.
\bibitem{ermushev} A.V. Ermushev,
D.I. Kazakov and O.V. Tara\-sov, \np{B281} {72} {87}
D.I. Kazakov, {\sl Phys. Lett.} {\bf B179} (1986) 952.
\bibitem{DJZ} X.D.~Jiang and X.J.~Zhou,
{\sl Phys. Rev. D} {\bf 42} (1990) 2109.
\bibitem{strassler}
M.J. Strassler, RU-95-86, hep-th/9602021,
and references therein.
\bibitem{LuZou} C. Lucchesi and G. Zoupanos,
Proc. of STU-Dualities Workshop,
CERN, Nov 27 - Dec 1, 1995, hep-ph/9604216.
\bibitem{pisi} O. Piguet and K. Sibold,
{\sl Int. J. Mod. Phys.}
{\bf A1} (1986) 913; {\sl Phys. Lett.}
{\bf B177}\,(1986)\,373.
\bibitem{HPS} S. Hamidi, J. Patera and
J.H. Schwarz,
{\sl Phys. Lett.} {\bf B141} (1984) 349.
\bibitem{LPnotes} C. Lucchesi and O. Piguet,
unpubl. notes.
\end{thebibliography}
\end{document}